\documentclass[fleqn,usenatbib]{mnras}

\usepackage{newtxtext,newtxmath}
\usepackage[T1]{fontenc}
\usepackage{float}
\usepackage{graphicx}
\usepackage[caption=false]{subfig}
\DeclareRobustCommand{\VAN}[3]{#2}
\let\VANthebibliography\thebibliography
\def\thebibliography{\DeclareRobustCommand{\VAN}[3]{##3}\VANthebibliography}
\usepackage{color}
\usepackage{ulem}

\usepackage{graphicx}	% Including figure files
\usepackage{amsmath}	% Advanced maths commands
\title[Tracing the cosmic web around galaxy clusters]{The probability of identifying the cosmic web environment of galaxies around clusters motivated by the Weave Wide Field Cluster Survey}

\author[D. Cornwell et al.]{\parbox{\textwidth}{
  Daniel J. Cornwell$^{1}$\thanks{E-mail: daniel.cornwell@nottingham.ac.uk}, 
  Alfonso Arag\'{o}n-Salamanca$^{1}$, 
  Ulrike Kuchner$^{1}$,
  Meghan E. Gray$^{1}$, 
  Frazer R. Pearce$^{1}$,
  Alexander Knebe$^{2,3,4}$
  } 
\\
\\
$^{1}$School of Physics and Astronomy, University of Nottingham, Nottingham NG7 2RD, UK\\
$^{2}$Departamento de F\'isica Te\'{o}rica, M\'{o}dulo 15, Facultad de Ciencias, Universidad Aut\'{o}noma de Madrid, 28049 Madrid, Spain\\
$^{3}$Centro de Investigaci\'{o}n Avanzada en F\'isica Fundamental (CIAFF), Facultad de Ciencias, Universidad Aut\'{o}noma de Madrid, 28049 Madrid, Spain\\
$^{4}$International Centre for Radio Astronomy Research, University of Western Australia, 35 Stirling Highway, Crawley, Western Australia 6009, Australia
}

\date{Accepted XXX. Received YYY; in original form ZZZ}

\pubyear{2023}

\begin{document}
\label{firstpage}
\pagerange{\pageref{firstpage}--\pageref{lastpage}}
\maketitle

\begin{abstract}
Upcoming wide-field spectroscopic surveys will observe galaxies in a range of cosmic web environments in and around galaxy clusters. In this paper, we test and quantify how successfully we will be able to identify the environment of individual galaxies in the vicinity of massive galaxy clusters, reaching out to $\sim5R_{200}$ into the clusters' infall region. We focus on the WEAVE Wide Field Cluster Survey (WWFCS), but the methods we develop can be easily generalised to any similar spectroscopic survey. Using numerical simulations of a large sample of massive galaxy clusters from \textsc{TheThreeHundred} project, we produce mock observations that take into account the selection effects and observational constraints imposed by the WWFCS. We then compare the `true' environment of each galaxy derived from the simulations (cluster core, filament, and neither core nor filament, {``NCF''}) with the one derived from the observational data, where only galaxy sky positions and spectroscopic redshifts will be available. We find that, while cluster core galaxy samples can be built with a high level of completeness and moderate contamination, the filament and NCF galaxy samples will be significantly contaminated and incomplete due to projection effects exacerbated by the galaxies' peculiar velocities.  We conclude that, in the infall regions surrounding massive galaxy clusters, associating galaxies with the correct cosmic web environment is highly uncertain. However, with large enough spectroscopic samples like the ones the WWFCS will provide (thousands of galaxies per cluster, {out to $5R_{200}$}), and the correct statistical treatment that takes into account the probabilities we provide here, we expect we will be able to extract robust and well-quantified conclusions on the relationship between galaxy properties and their environment.

\end{abstract}


%%%%%%%%%%%%%%%%%%%%%%%%%%%%%%%%%%%%%%%%%%%%%%%%%%

%%%%%%%%%%%%%%%%% BODY OF PAPER %%%%%%%%%%%%%%%%%%

\section{Introduction}
\label{sec:intro}
Galaxies are not distributed through the Universe randomly. Instead, they are constituents of the larger-scale cosmic web. This network features nodes -- peaks in the matter density field, as well as sheets, filaments, walls, and voids \citep{aragon10}. First described in \cite{Zeldovich}, these structures are ubiquitous and form from the anisotropic gravitational collapse of dark matter. 

As galaxies assemble their mass and evolve, they may experience a range of environments which can significantly influence their properties. Most notably, galaxies in the densest cosmic environments, (i.e., galaxy clusters), frequently interact with other galaxies, the intracluster medium (ICM), and the cluster tidal field. During their lifetimes, galaxies may also travel through cosmic sheets and filaments, where their properties can be affected long before they reach the clusters themselves. This is often referred to as "pre-processing" \citep{Zabludoff_1998}. Processing and pre-processing can lead to strong environmental differences in the properties of galaxies, such as the Morphology--Density relation {\citep{Dressler_80}} {that describes the finding of an increased} fraction of early-type galaxies in denser environments.

Since galaxy clusters are part of the cosmic web, they co-evolve with it and with the galaxies that live in them. As clusters grow through galaxy infall and mergers with other clusters and groups, their dynamical state changes, and this can also influence the environmental effects their galaxies experience \citep{De_Luca21, Ribeiro13, Morell20}. Galaxy groups and their central galaxies can also be significantly affected by the presence of filamentary structures that feed into them \citep{Poudel_2017}. It is therefore clear that a complete understanding of how the environment affects galaxy evolution requires a thorough mapping and study of the cosmic web around galaxy clusters. 

The importance of clusters and filaments is emphasised by the fact that, even though filaments and clusters only contain $6\%$ and $0.1\%$ respectively of the volume of the present-day universe, they harbour $50\%$ and $11\%$ of the mass \citep{Cautun14}. Since the physical processes galaxies experience depend on environment, it is vital not only to map these environments accurately, {but also to be able to find well-defined subsets of galaxies in each environment.} 

The identification of cosmic filaments has been rigorously tested {and applied} in large-area surveys where a variety of detection methods have been adopted, such as the widely used discrete persistent structure extractor tool \texttt{DisPerSE} \citep{Sousbie11, Sousbie_2_11}. \texttt{DisPerSE} has been successfully applied to optical surveys such as SDSS \citep{Malavasi20} and GAMA \citep{Kraljic18}. Cosmic filaments are also detectable in X-rays \citep{vernstrom2021discovery} and follow-up X-ray studies of filaments found in the SDSS has resulted in a $5\sigma$ detection of their X-ray emission \citep{Tanimura_2022}. {Several} other geometrical web extractors exist that use alternative methods (see \cite{Libeskind} for a review on cosmic web tracing algorithms). Whilst these cosmic web finders are effective at mapping these environments over scales of hundreds of Mpc, mostly far from galaxy clusters, {observationally} characterizing the environment in the vicinity of massive clusters {remains a challenge.} %is a considerably more difficult task. 
This is due primarily to the complexity of the infall regions around clusters -- where multiple filaments converge -- and the large peculiar velocities induced by the cluster dynamics {\citep{Tempel_2016,Kuchner21}}. It is therefore clear that mapping the cosmic web in the vicinity of clusters requires special attention. 

This will be addressed by next-generation wide-field spectroscopic surveys covering regions of tens of Mpc around galaxy clusters. Examples of such surveys include the WEAVE Wide Field cluster Survey  \citep[WWFCS;][Kucher et al. in prep]{Jin} and the 4MOST CHileAN Cluster galaxy Evolution Survey ({CHANCES}, Haines et al.\ in prep). This paper focuses on the first of these survey, which is an upcoming wide-field spectroscopic study of $\sim 20$ galaxy clusters to be undertaken at the $4.2\,$m William Hershel telescope with the newly-commissioned WEAVE spectrograph \citep{Jin}. The WWCS will obtain thousands of spectra for galaxies in and around each cluster, reaching out to several virial radii from the cluster cores. This will allow a thorough investigation of the cosmic web around galaxy clusters and the properties of the galaxies in it.

To fully exploit the extensive data generated by the WWFCS, we are developing specific analysis techniques to test and optimize the detection and characterisation of the filamentary networks around the clusters \citep{Kuchner20, Kuchner21, Kuchner22, Cornwell_2022} using simulated clusters from 
 \textsc{TheThreeHundred} project \citep[][see below for more details]{Cui2018, Cui22}. Specifically, \cite{Kuchner21} investigated the practicalities of extracting filament networks around the clusters in the presence of the observed {``Fingers of God'' (FoG)} due to the galaxies' peculiar velocities. They concluded that, because the distance uncertainties induced by these peculiar velocities are comparable with the depth of the volume explored, filament extraction near galaxy clusters need to rely on two-dimensional projections on the sky. This does not mean that the spectroscopic redshifts are not necessary -- they are crucial to reliably select the galaxies that belong to the relevant volume around the cluster with an accuracy of $\sim10\,$Mpc. Such accuracy cannot be achieved with photometric redshifts.  

Bringing the \textsc{TheThreeHundred} simulations one step closer to observations, \cite{Cornwell_2022} demonstrated that filaments can be successfully extracted from datasets similar to those expected from the WWFCS. They did that by creating mock observations from the simulated galaxy samples and considering the effects of sample selection and completeness resulting from the spectrograph's fibre allocation and spatial coverage.  They found that the filament networks extracted in the simulations are well reconstructed in the mock observations. In this paper, we go a{n additional} step forward and quantify our ability to allocate galaxies to the different environments around the { simulated mock-observed} clusters. The resulting statistics will be necessary when studying the properties of galaxies as a function of environment in a robust statistical way.  

This paper is structured as follows. Section~2 describes the (expected) observational and simulated datasets. Section~3 presents the reference filament networks derived from the simulated galaxy samples and defines the `true' environment of each galaxy. Section~4 describes how the filaments are found in the mock WWFCS-like observations, and how the `observed' environment of each galaxy is determined. Section~5 compares the `true' and `observed' environments of the galaxies, and provides the necessary statistics to quantify the success (or otherwise) of the comparison. Section~6 summarises the conclusions.

\section{Dataset}
\subsection{WEAVE Wide Field Cluster Survey}
\label{sec:WWFCS}
WEAVE (the William Herschel Telescope Enhanced Area Velocity Explorer) is a new optical multi-object spectrograph on the $4.2\,$m William Herschel Telescope \citep{Balcells10, Dalton14, Dalton16, Jin}. The WEAVE Wide-Field Cluster Survey is one of the surveys that will be carried out with WEAVE. It is  designed to yield thousands of galaxy spectra in and around galaxy clusters (Kuchner et al., in prep.). The WWFCS will take advantage of the multiplex capability provided by $\sim1000$ individual fibres deployable over a 2-degree-diameter field-of-view to observe up to 20 low-redshift ($0.04 < z < 0.07$) galaxy clusters. {The \textit{r}-band magnitude limit of the WWFCS spectroscopic observations is 19.75, corresponding to an approximate galaxy stellar mass limit of $10^{9} M_{\odot}$}. The WWFCS will use the low spectral resolution mode ($R\sim 5000$) to obtain optical spectra in the $366 \text{nm} < \lambda < 959\text{nm}$ range. For each cluster we will obtain spectra for several thousands of potential cluster members, reaching beyond $\sim 5R_{200}$ from the cluster centre. These spectra will not only provide accurate redshifts, but also yield key information on the stellar populations and AGN properties of the galaxies. 

Even though each WEAVE field can target close to one thousand objects in a single pointing, and we will observe each cluster with several pointings, it is not feasible to target every single galaxy in the cluster region due {to instrumental limitations that include} potential fibre collisions and overlap. WEAVE uses an algorithm named \texttt{Configure} \citep{Terret14} to optimise fibre allocation. It was shown by \cite{Cornwell_2022} that \texttt{Configure} is able to allocate fibres to $\sim 75\%$ of the target galaxies overall, reaching $81\%$ outside $R_{200}$. As shown by these authors, this is enough to reliably reconstruct the filament networks around the clusters. {In the current paper, we base our analysis on mock-observed (i.e., ``configured'') simulated galaxy samples in clusters and their outskirts.}

\subsection{TheThreeHundred simulations of galaxy clusters}
\label{sec:The300}
\textsc{TheThreeHundred}\footnote{\url{https://the300-project.org/}} project \citep{Cui2018} is a set of zoom-in resimulations of the Multidark Dark Matter only (MDPL2) cosmological simulations \citep{Klypin2016}. MDPL2 is a periodic cube of comoving length $1\,h^{-1}\,$Gpc containing $3840^3$ dark matter particles, each with mass $1.5 \times 10^{9}\,h^{-1} {M_{\odot}}$. MDPL2 uses \textit{Planck} cosmology {($\Omega_{\text{M}} = 0.307,  \Omega_{\text{B}} = 0.048,  \Omega_{\Lambda} = 0.693,  h = 0.678,  \sigma_8 = 0.823,  n_s = 0.96$; \citealt{Planck15}}). 

This project locates the 324 most massive haloes ($M_{\text{200}} > 8 \times 10^{14} h^{-1}M_{\odot}$) at $z=0$, it then follows them back to their initial conditions, and resimulates the hydrodynamics in a 15 $h^{-1}$Mpc comoving sphere around the identified massive halo. The highest resolution dark matter particles are then split into dark matter and gas, following the cosmological baryonic mass fraction using the Planck 2015 cosmology $m_{\text{DM}} + m_{\text{gas}} = 1.5 \times 10^{9} h^{-1} M_{\odot}$. We use the outputs of the zoom re-simulations ran using \textsc{Gadget-X}, which incorporates full-physics galaxy formation, star formation, and feedback from both SNe and AGN.  The work in this paper utilizes the AMIGA Halo Finder \citep{Gill2004,Knollmann2009} to determine the halo properties and we take the redshift $z=0$ snapshot which is comparable to the low redshifts of the WWFCS clusters. For the purpose of this work, we take the galaxy's stellar mass to be one tenth of the dark matter halo mass, as justified in \cite{Kuchner20}. Our team has successfully exploited the \textsc{TheThreeHundred} simulations to plan the WWFCS observations and make predictions about the properties of the observed galaxy samples \citep{Kuchner20,Kuchner21,Cornwell_2022}, demonstrating that reliable mock observations that mimic the WWFCS can be generated from this suite of simulations. Specifically, in \cite{Cornwell_2022} we mass-matched each of the 16 main clusters targeted by the WWFCS with 10 clusters in \texttt{TheThreeHundred} simulations, and mock WWFCS observation were created for all 160 of them. {{As explained above, we use \texttt{Configure} to mock-observe the galaxies in and around the mass-matched clusters. G}oing forward, we refer to these cluster galaxy samples as the 2D mock-observations. Throughout this paper we will also use the galaxies in the full 3D simulated clusters {before ``configurations''} and refer to these as the ``true simulated cluster galaxy samples''.}  

In the next section we outline how we identify the ``true'' cosmic web structures in the simulations that we will later compare with the ones extracted from the mock observations.

\section{Defining cosmic web environments}
\label{sec:filaments}
\subsection{Cosmic web {extraction}}
\label{sec:Cosmic_filaments}
Following \cite{Cornwell_2022}, we employ the widely used structures extractor algorithm \texttt{DisPerSE} \citep{Sousbie11, Sousbie_2_11} to identify filaments in the simulation boxes at $z = 0$. To match the expected depth of the WWFCS, we use as `mock galaxies' all the halos in the simulations with a dark matter mass larger than $10^{10}M_\odot$, approximately corresponding to stellar masses larger than $10^{9}\,M_\odot$ \footnote{We note that the halo and stellar mass limits are slightly higher for more massive clusters, as described in \cite{Cornwell_2022}}. \texttt{DisPerSE} identifies persistent topological features in the underlying density field defined by the mock galaxies. These features include {critical points such as peaks and segments that link the peaks along the local geometry of ridges, which are the} filaments. In \cite{Cornwell_2022} we explain in detail how we use \texttt{DisPerSE} to extract robust filamentary networks around the simulated clusters, including our choice of input parameters. Following \cite{Kuchner20}, we use mass-weighting since this option makes the filamentary networks more reliable. This choice has some observational support from the fact that galaxies in filaments tend to be more massive (as well as redder and less star-forming) than field galaxies away from them  \citep{Kraljic18,Chen17, Malavasi17, Laigle18, Florian}.

\subsubsection{3D reference filament networks in simulations}
\label{sec:3d_nets}
In order to get consistent and reliable `true' filament networks from the simulations we need to choose some critical input parameters for \texttt{DisPerSE}. Building on the work of \cite{Cornwell_2022} and Cornwell et al.\ (in prep.), we use a \textit{persistence} value of $4.6\sigma$ and a \textit{smoothing} of 5 (see \citealt {Sousbie11} and \citealt{Sousbie_2_11} for a definition of these parameters). {With the persistence ratio threshold, we vary the robustness or significance of filaments to local variations in the density field. In our case, we intend to extract the primary filaments that are responsible for the majority of cluster accretion. Note that the use of} mass-weighting {requires} a higher \textit{persistence} value to obtain networks that are similar to non-mass-weighted networks obtained with a lower \textit{persistence}. {The second parameter, \textit{smoothing}, creates smoother networks by averaging the position of each vertex with that of its direct neighbors.}

\subsubsection{2D mock-observational filament networks}
\label{sec:2Dfilaments}
We also need to define filament networks using a mock-observational dataset similar to the one WEAVE will provide \citep[cf.][]{Cornwell_2022}. The main two differences from the original simulations are that we will not observe every single galaxy (see Section~\ref{sec:WWFCS}) and that we will not have radial distances but only redshifts (or radial velocities). As explained in \cite{Cornwell_2022}, despite these limitations, reliable 2D filamentary networks (projected on the plane of the sky) can be extracted from the simulated datasets. Again, we use the mass-weighted \texttt{DisPerSE} algorithm to identify the components of the cosmic web in 2D and chose a \textit{persistence} of $2.6\sigma$ and a \textit{smoothing} of 5. 

\begin{figure*}
    \centering
    \includegraphics[width = \textwidth]{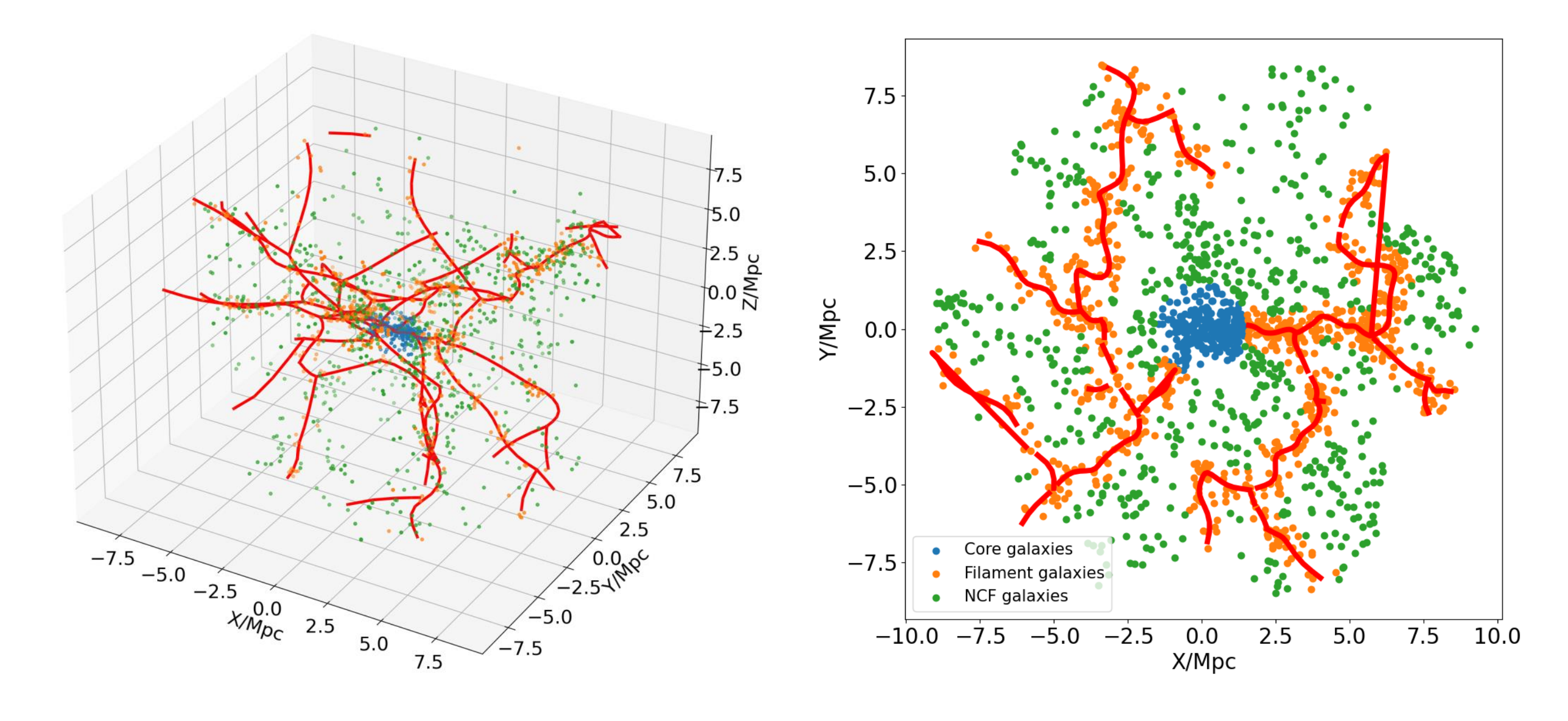}
    \caption{Illustration of an example cluster with the different cosmic web environments identified in 3D (directly from the simulations) and in 2D (from the mock observations based on WWFCS observational constraints). The environment assigned to a given galaxy may not be the same in 2D and 3D due to projection effects. Cluster core galaxies are shown in blue, filament galaxies in orange, and the rest of the galaxies (neither core nor filament, NCF) in green. We also overlay the associated filament network around this cluster, shown by the red lines. This simulated cluster has similar mass to RX0058, one of the targets in the middle of the mass range of the WWFCS sample.} %See Sections~\ref{sec:filaments} and~\ref{sec:assign_environments} for a description of how the filaments were identified and how galaxies are allocated to each environment.}
    \label{fig:example_environments}
\end{figure*}

The 3D and 2D filament networks around a simulated {example} cluster are shown in Figure~\ref{fig:example_environments}. 

\subsection{Filament thickness}
\label{sec:thickness}
One important issue to consider when allocating galaxies to filaments is the thickness of the filaments themselves. In other words, how close to the spine of a filament does a galaxy have to be in order to be considered a `filament galaxy'? 
This is not trivial since filaments do not have sharp boundaries. Some studies show that the thickness of filaments may depend on their length: longer filaments may be thinner, on average, than shorter ones \citep{refId0}.  Additionally, \cite{Rost21} used the gas and dark matter distributions from \textsc{TheThreeHundred} to suggest that filaments are the thickest {closest to} the nodes. \cite{Kuchner20} used the transverse gas density profile of filaments in the same simulations and derived a characteristic filament radius {$\sim0.7$--$1\,h^{-1}\,$Mpc} for massive clusters. 
However, there is no theoretical or observational motivation for holding the filament thickness constant for the networks surrounding different clusters. The mass range of the clusters selected for the WWFCS spans more than an order of magnitude (\citealt{Jin, Cornwell_2022}; Kuchner et al.\ in prep.). If the thickness of the filaments surrounding clusters with different masses is kept constant, the fraction of the total volume occupied by filaments inside a sphere of radius $\sim5R_{200}$ would be significantly larger for less massive clusters.  Self-similarity considerations on the dark matter distributions suggest that this should not be the case. In the absence of stronger evidence, it is reasonable to assume that the thickness of filaments surrounding a given cluster scales with the cluster's $R_{200}$, and therefore with $M_{\text{cluster}}^{1/3}$. Hence, 
\begin{equation}
    R_{\text{fil}} = R_{0} * \left( \frac{M_{\text{cluster}}}{\langle{M_{\textsc{TheThreeHundred}}}\rangle}\right)^{1/3},
    \label{eq:thickness}
\end{equation}
where $R_{0} = 0.7\,h^{-1}\,$Mpc is the average filament radius calculated using the full \textsc{TheThreeHundred} cluster simulations \citep{Kuchner20}, $\langle{M_{\textsc{TheThreeHundred}}}\rangle$ is the average cluster mass in the \textsc{TheThreeHundred} sample ($6.5\times10^{14}\,M_\odot$), and $M_{\text{cluster}}$ is the mass of the cluster at the centre of a specific filament network. This thickness is kept constant for all filaments surrounding this particular cluster. {As discussed above}, this is an over-simplification, but alternative recipes could be easily applied. We keep this one for simplicity throughout the paper.

\section{Assigning galaxies to cosmic web environments}
\label{sec:assign_environments}
In this section, we provide a framework for assigning mock galaxies to the different cosmic web environments associated with galaxy clusters. This is an essential but non-trivial process if we wish to understand how each environment affects the galaxies' properties and evolution. 

Simulations provide full 3D positional information, allowing a precise `true' environmental assignments to each galaxy. Observations will provide accurate sky positions and radial velocities, but the radial distances will have uncertainties that are large in comparison with the physical dimensions of the volume considered (a sphere with a radius $\sim5R_{200}$), meaning that the environmental assignments from the observations can only be done in 2D (see Section~\ref{sec:2Dfilaments}). This means that the environment associated to a galaxy from the observational data may not coincide with the `true' environment. We need to quantify statistically how often that occurs.

We assign each galaxy to different environments according to the following criteria: 
\begin{enumerate}
  \item Cluster core galaxy: galaxies that lie at a radial distance from the cluster centre $r < R_{200}$ in 3D or 2D. The blue points in Figure~\ref{fig:example_environments} show these galaxies.
  \item Filament galaxy: galaxies that lie outside the cluster core ($r > R_{200}$) and {close to the filament spine} ($r_{\text{spine}} < R_{\text{fil}}$), where $r_{\text{spine}}$ is the perpendicular distance from the spine of the filament and $R_{\text{fil}}$ is the filament thickness determined using Equation~\ref{eq:thickness}. These galaxies are represented by the orange points, whilst the filaments' spines are shown as red lines in the figure.
  \item NCF galaxy (neither core nor filament): galaxies that lie outside of the cluster core and outside of filaments ($r > R_{200}$ and $r_{\text{spine}} > R_{\text{fil}}$). They are shown as green points.\footnote{We note that is not an exhaustive list of cluster substructure and we focus in a follow up paper (Cornwell et al. in prep,) on the detection of galaxy groups in mock-observations.} 
\end{enumerate}

\begin{table*}
\caption{Fraction of galaxies in each environment and fraction of the total volume/surface area occupied by each environment within $r < 5R_{200}$, 
 in 3D and 2D.}
\label{tab:environments}
\begin{tabular}{lcccc}
\hline
Environment & 3D population & 3D volume & 2D population & 2D surface area\\
\hline
Cluster & 10\% & 1\% & 15\% & 4\% \\
Filament & 38\% & 6\% & 45\% & 19\%\\
NCF & 52\% & 93\% & 40\% & 77\%\\

\hline
\end{tabular}
\end{table*}

The proportion of galaxies identified in each environment, together with the fraction of the total volume/surface area that each environment occupies around the simulated galaxy clusters in 3D and 2D are presented in Table~\ref{tab:environments}. Not surprisingly, we find that the densest environments are the cluster cores, followed by filaments, with the NCF region being much less dense. Within $5R_{200}$, the cluster cores contain $\sim10\%$ of the galaxies, while they occupy only $1\%$ of the volume, and filaments contain $\sim36\%$ of the galaxies in $\sim6$\% of the volume. As a comparison, over much larger regions ($\sim500\,$Mpc scales), \cite{Cautun14} found that filaments contain $\sim6$\% of the volume, whilst accounting for half of the total mass budget. Similarly, they also found that nodes, which are proxies for cluster cores \citep{Cohn22}, contain $\sim10$\% of the mass but only $0.1\%$ of the volume. Table~\ref{tab:environments} also shows that the density contrast between the different environments is significantly reduced when we move from 3D to 2D due to projection effects.

\section{Results}

In what follows we will quantify statistically how often the `true' environment assigned to a galaxy in 3D (directly from the simulations) agrees with the one identified in 2D, once the limitations imposed by observation such as the WWFCS are taken into account {(see Sec. \ref{sec:WWFCS})}. This information is essential in order to interpret the observational data correctly when trying to infer how different environments affect galaxy properties and evolution.

\subsection{Overall performance of environment allocation}
\begin{figure*}
    \centering
    \includegraphics[width = \textwidth]{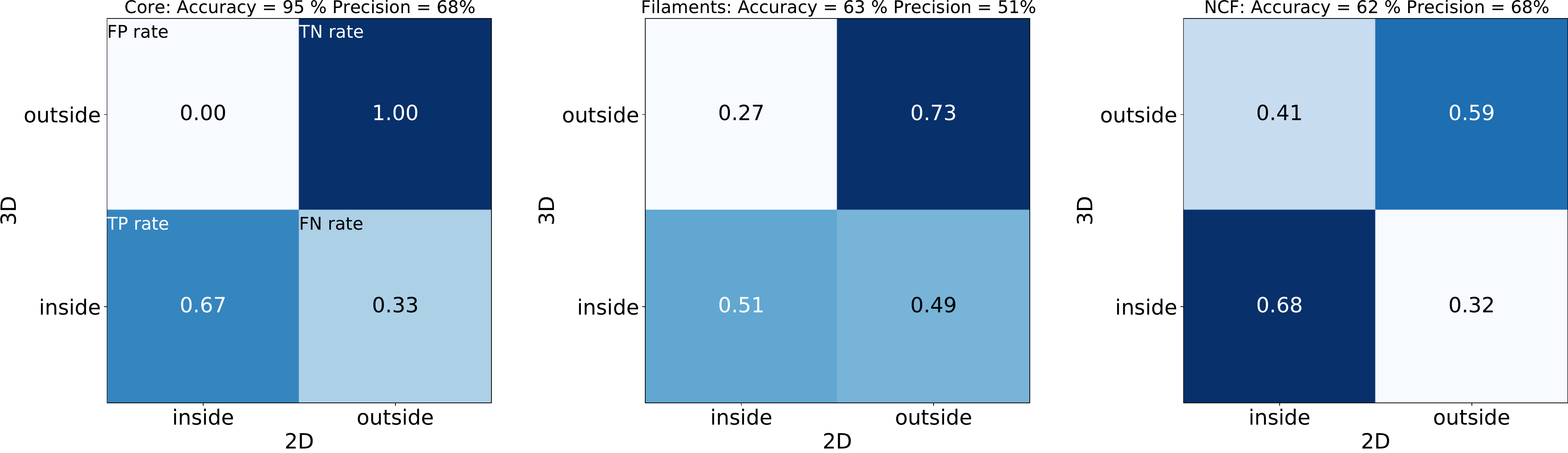}
    \caption{Confusion matrices comparing the environment identification for galaxies in 3D (models) and 2D (simulated {mock} observations). In each panel, the top left box represents the false positive rate, the bottom left box is the true positive rate, the top right box is the true negative rate, and the bottom right is the false negative rate. Left panel: galaxies identified as core galaxies. Middle panel: galaxies identified as filament galaxies. Right panel: galaxies identified as neither core nor filament galaxies (NCF). The accuracy (Equation~\ref{eq:accuracy}) and precision (Equation~\ref{eq:precission}) are displayed at the top of each panel.}
    \label{fig:confuse_all}
\end{figure*}

In Figure~\ref{fig:confuse_all}, we display three confusion matrices to assess how well we can allocate galaxies to different environments using WWFCS-like data. A confusion matrix is a way of visualizing the success of binary classification. In our case, we allocate galaxies to different environment using the mock observations (2D) and compare them to the `true' environment determined from the full 3D simulations, as described in previous sections. We use the following {standard} definitions of true-positive, true-negative, false-positive and false-negative for environment X (where X can be core, filament or NCF):
\begin{enumerate}
     \item[$TP=$] number of true-positives $=$ number of galaxies identified as belonging to environment X in the mock observations (2D) and to the same environment X in the simulations (3D). 
    \item[$TN=$] number of true-negatives $=$ number of galaxies identified as belonging to environment Y ($\neq\,$X) in the mock observations (2D) and to environment Y in the simulations (3D).
    \item[$FP=$] number of false-positives $=$ number of galaxies identified as belonging to environment X in the mock observations (2D) and to environment Y in the simulations (3D).
    \item[$FN=$] number of false-negatives $=$ number of galaxies identified as belonging to environment Y in the observations (3D) and to environment X in the simulations (3D). 
\end{enumerate}

Furthermore, we define accuracy as 
\begin{equation}
\label{eq:accuracy}
Accuracy =  \frac{TN+TP}{TP+TN+FP+FN},   
\end{equation}
and precision as
\begin{equation}
\label{eq:precission}
Precision = \frac{TP}{TP+FP}.
\end{equation}

In the left panel of Figure~\ref{fig:confuse_all} we illustrate the success of classifying cluster core galaxies. For this environment, we get an accuracy of $95\%$ and a precision of $68\%$. The true-positive identification rate (fraction of galaxies correctly identified in 2D as belonging to the cluster core) is $67$\%. Because of projection effects, about one third of the galaxies identified as belonging to the core in 2D are either in front or behind the core itself. For obvious reasons, the `outside' row of the confusion matrix shows that all galaxies that are outside the core in 2D are also outside in 3D. 

The second and third panels evaluate our success (or lack thereof) at identifying filament and NCF galaxies in the 2D mock observations. The true-positive rate for filament galaxies (fraction of galaxies correctly identified in 2D as belonging to filaments) is relatively small ($51$\%), {while we are able to correctly identify galaxies not belonging to filaments in $73$\% of the case (true-negative rate}. Conversely, we have a higher success at identifying NCF galaxies (true-positive rate of $68$\%) than at rejecting them (true-negative rate of $59$\%). This can be easily understood by considering projection effects and the fraction of the total surface area covered by each one of these environments. 

These results indicate that identifying the correct environment of a galaxy in the vicinity of a cluster is {not straightforward}, and the resulting statistical uncertainties cannot be ignored when interpreting the observations. We will investigate next how these uncertainties depend on the distance of a galaxy to the cluster centre and its mass.

\subsection{The dependence of the environmental identification success on galaxy mass and clustercentric distance}
 In this section we will quantify our success at assigning environments to galaxies using the information provided by spectroscopic surveys like the WWFCS. The goal is to be able to answer the question: if we assign a given environment to a galaxy based on the observations, what is the probability that it is truly in that environment? And, importantly, how does this change with a galaxy's position and mass?

\subsubsection{Evaluating a single cluster}
\label{sec:evaluating_single_cluster}

\begin{figure*}

\subfloat{
  \includegraphics[clip,width=0.99\textwidth]{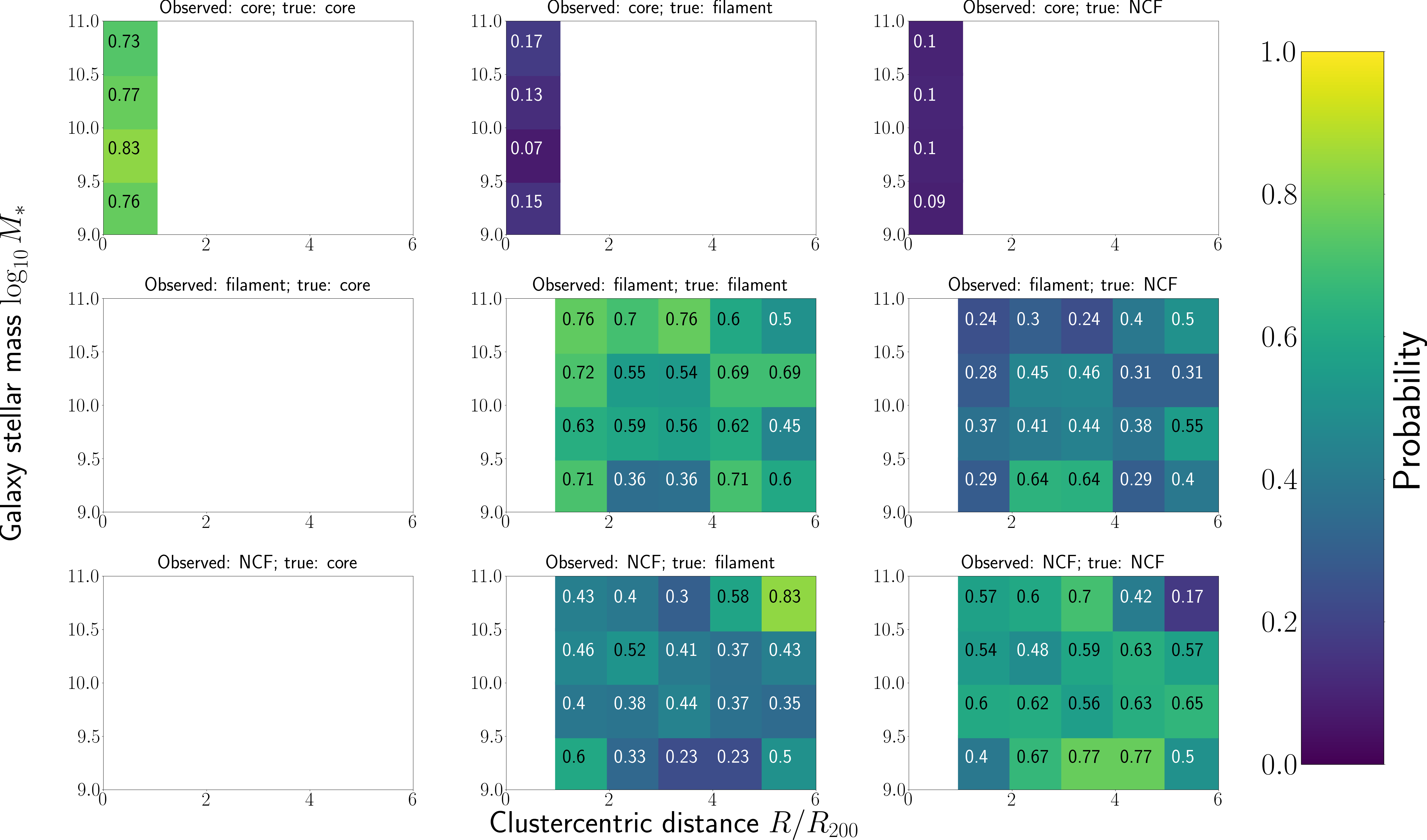}%
}

\subfloat{
  \includegraphics[clip,width=0.99\textwidth]{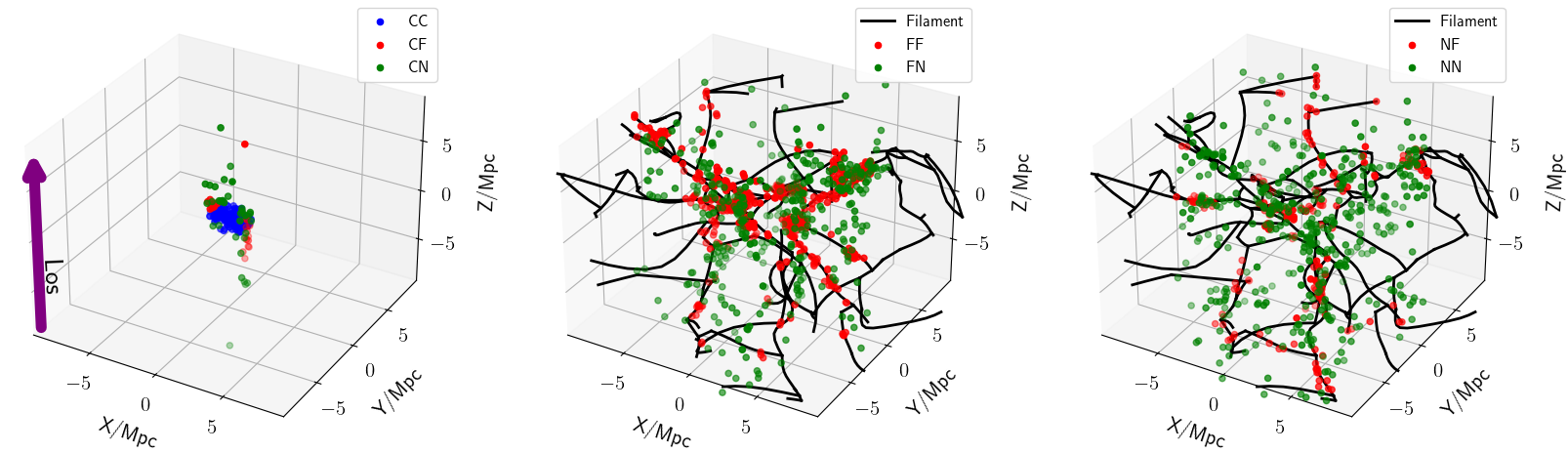}%
}
    \caption{The first 9 panels ($3\times3$ grid) display the probabilities of identifying galaxies in different environments (cluster core, cosmic filaments, or neither core nor filament, NCF)) as a function of galaxy stellar mass and clustercentric distance {in one example cluster}. For every bin, we calculate the probability of correct identification  $P_{\text{XX}ij}$ (panels in the top-right to bottom-left diagonal) as well as incorrect identifications  $P_{\text{XX}ij}$ (other panels) using Equations~\ref{eq:probxx} and~\ref{eq:probxy}. For illustration we show this for one model cluster, the same one shown in Figure~\ref{fig:example_environments}. In the bottom row we show the spatial distribution of galaxies in the different environments for the same model cluster. The left panel in the bottom row shows all the galaxies that are identified as core galaxies in 2D, with the ones correctly identified as belonging to the core in blue, those misidentified as filament galaxies in red, and the ones misidentified as NCF in green. Similarly, the middle panel of the bottom row shows all the galaxies identified as filament galaxies in 2D, with the ones correctly identified as filament galaxies in red, and those misidentified as NCF galaxies in green. The filament network is shown as black lines. Finally, the right panel in the bottom row shows galaxies that are identified as NCF galaxies in 2D, with the ones correctly identified NCF galaxies in green and those misidentified as filament galaxies in red. The line-of-sight of the cluster is indicated by the purple arrow, parallel to the $z$ axis. }
    \label{fig:truth_table_example}
\end{figure*}

We start by using a typical cluster to describe the process. We choose a simulated cluster with a mass similar to WWFCS cluster RX0058, the cluster shown in Figure~\ref{fig:example_environments}. We first divide all the galaxies in mass and 2D clustercentric radial distance bins. For all the galaxies in bin $(i,j)$, where $i$ corresponds to a mass bin and $j$ to a radial distance bin, we calculate the probability $P_{\text{XX}ij}$ that a galaxy has been \textit{correctly} allocated to environment X in the mock observations (2D) as the ratio of the number of galaxies allocated to environment X in 2D whose `true' 3D environment is also X, $N_{\text{XX}ij}$, and the total number of galaxies allocated to environment X in 2D $N_{\text{X}ij}$. In other words,
\begin{equation}
\label{eq:probxx}
   P_{\text{XX}ij} = \frac{N_{\text{XX}ij}}{N_{\text{X}ij}}.
\end{equation}
Conversely, we calculate the probability $P_{\text{XY}ij}$ that a galaxy has been \textit{incorrectly} allocated to environment X in 2D when its `true' 3D environment is Y ($\neq$X) as the ratio of the number of galaxies allocated to environment X in 2D whose `true' 3D environment is Y, $N_{\text{XY}ij}$, and the total number of galaxies allocated to environment X in 2D $N_{\text{X}ij}$. Hence,
\begin{equation}
\label{eq:probxy}
   P_{\text{XY}ij} = \frac{N_{\text{XY}ij}}{N_{\text{X}ij}}.
\end{equation}
We calculate this probability in four galaxy stellar mass bins and five clustercentric distance bins, covering the ranges $10^{9} M_{\odot} < M_{*} <10^{11} M_{\odot}$ and $0 < r < 6R_{200}$ ({note that all cluster regions are fully covered out to $5R_{200}$ by the WWFCS pointings, and in some cases we reach beyond $6R_{200}$}). The resulting probabilities in each mass and radial bins for the chosen cluster are shown in Figure~\ref{fig:truth_table_example}. 

The first row of this figure displays the probability of identifying a galaxy as belonging to the cluster core in 2D when its true 3D environment is the core, a filament or NCF (left, centre and right panels). For example, if we look in the $0<r<1R_{200}$ and $10^{9} M_{\odot} < M_{*} < 10^{9.5} M_{\odot}$ bin in the first row, we assign a probability of 0.76 for correctly identifying core galaxies as such, a probability of 0.15 for misidentifying a core galaxy as a filament galaxy, and a probability of 0.09 for misidentifying a core galaxy as an NCF galaxy. Reassuringly, the probability of correctly identifying a core galaxy is the highest by a large margin -- although there is some contamination due to projection effects, identifying core galaxies is relatively easy. This is illustrated in the bottom left panel of  Figure~\ref{fig:truth_table_example}, where `true' core galaxies correctly identified in 2D as belonging to the core are plotted as blue dots, core galaxies misidentified as belonging to filaments in red, and those misidentified as NCF in green. The incorrect identifications are purely a product of the contamination of the true cluster core sample due to projecting the 3D galaxy distribution in 2D.  

The second row of Figure~\ref{fig:truth_table_example} shows the probability of identifying a galaxy as belonging to a filament in 2D when in 3D it is a core galaxy, a filament galaxy, or an NCF galaxy (left, centre, and right panels respectively). The left panel is blank because it is not possible to identify a true core galaxy as a filament galaxy in 2D since in our framework the 2D filament networks inside a circle with projected radius of $R_{200}$ {are not taken into account} (see \citealt{Cornwell_2022}). The middle panel shows the probabilities of correctly classifying filament galaxies, while the right panel present the probability of misidentifying them as NCF galaxies. Notwithstanding the statistical fluctuations, the likelihood of a correct identification for filament galaxies is generally higher than that of a misidentification, but not by much (see Section~\ref{sec:evaluating_whole_sample}). The middle panel of the bottom row in Figure~\ref{fig:truth_table_example} shows the spatial distribution of the `true' filament galaxies correctly identified in 2D as belonging to filaments (red dots), and the filament galaxies misidentified in 2D as NCF (red dots). The 3D filament network is shown as black lines. 

In a similar way, the third row of Figure~\ref{fig:truth_table_example} presents the probability of classifying a galaxy in 2D as an NCF galaxy when it is truly in the cluster core, in a filament, or correctly identified as an NCF galaxy. Again, the left panel is blank because, by {our} definition, a true core galaxy cannot be classified as NCF. As before, the likelihood of correct identification of NCF galaxies is generally a bit higher than that of misidentification.  The right panel of the bottom row in this figure shows the spatial distribution of the `true' NCF galaxies correctly identified in 2D as NCF (green dots), and the NCF galaxies misidentified in 2D as filament galaxies (green dots). The 3D filament network is also shown as black lines.

\begin{figure*}
    \centering
    \includegraphics[width = \textwidth]{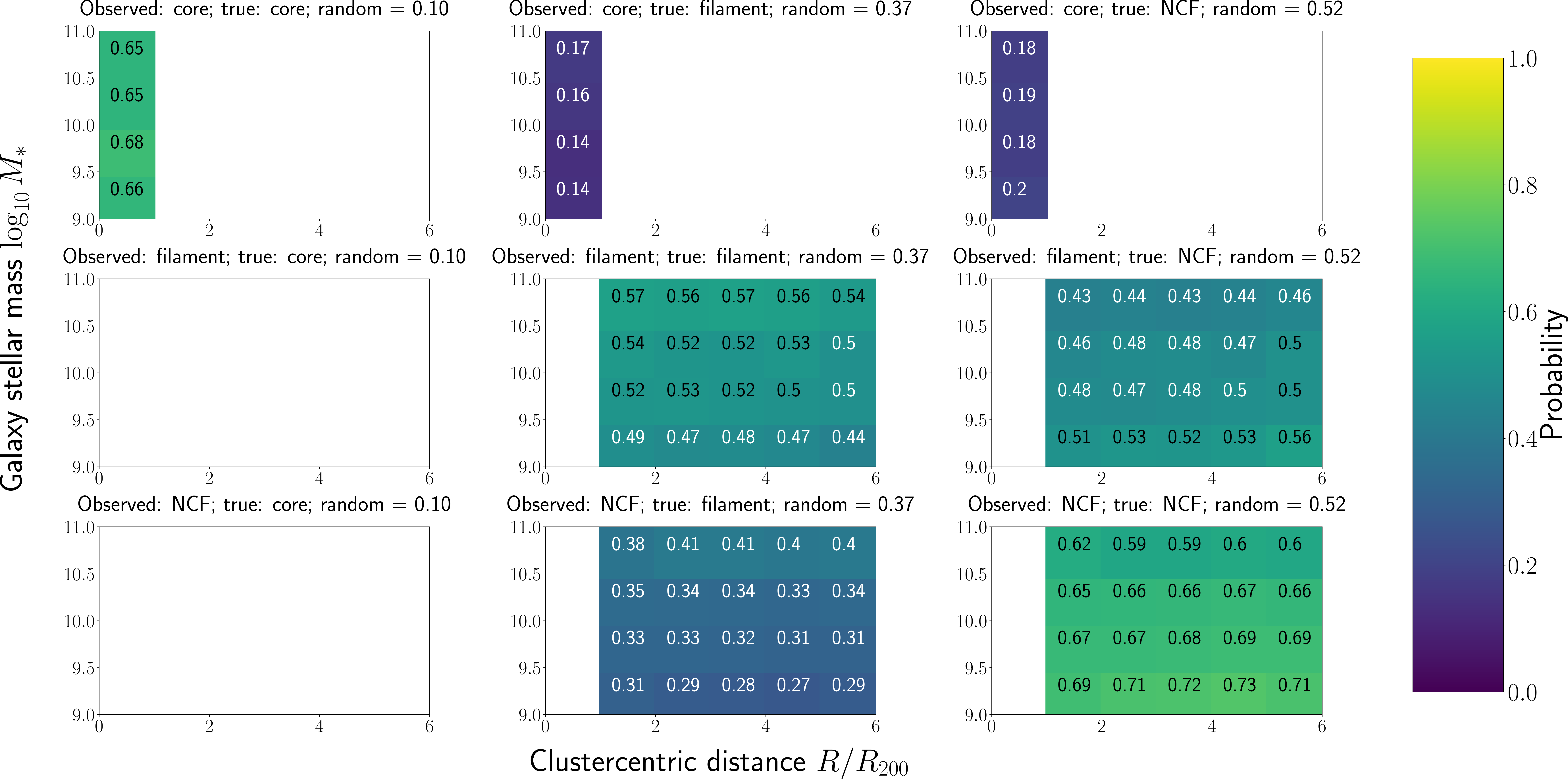}
    \caption{The same as the top $3\times3$ panels of Figure~\ref{fig:truth_table_example} but averaged over all clusters. The probability of randomly classifying a galaxy correctly in the title of each plot (see text for details). The information in this Figure is also included in Table \ref{tab:all}. }
    \label{fig:truth_table_all}
\end{figure*}

\subsubsection{Evaluating the entire simulated cluster sample}
\label{sec:evaluating_whole_sample}

We repeat the process described in Section~\ref{sec:evaluating_single_cluster} for the whole sample of 160 simulated clusters presented in \cite{Cornwell_2022}, see Section~\ref{sec:The300}, and compute the classification probabilities for each galaxy mass and radial bin in all the simulated clusters. These probabilities are then averaged in order to improve the statistics, and the standard deviation for each bin calculated. With this information we are now able to evaluate robustly any possible dependence on galaxy mass and radial position. Moreover, with such a large simulated cluster sample we will explore cluster-to-cluster variations in Section~\ref{sec:cluster_mass_dependence}. The average probabilities for each mass and radial bin are displayed in Figure~\ref{fig:truth_table_all} using the same format as Figure~\ref{fig:truth_table_example}, with the corresponding standard deviations shown in Figure~\ref{fig:truth_table_SD}.

 When assessing our ability to identify the environment of a galaxy correctly, we need to compare it with that of a random allocation; after all, if all our machinery does not perform better than random, no statistical inference will be possible. We calculate the probability of randomly allocating an environment to a galaxy by shuffling the environment labels for all galaxies {in the 3D simulations} and re-calculating the probabilities for all clusters individually, and then averaging them. {In doing so, we are using the 3D population statistics from Table \ref{tab:environments} as a prior in the random allocation since, in the absence of any other information, this is our `best guess' distribution of galaxies in each cosmic web environment.} The average random probabilities are displayed in the titles of each panel of Figure~\ref{fig:truth_table_all} for easy reference. Reassuringly, the random probabilities thus calculated are very close to the true fractions shown in Table~\ref{tab:environments}, as expected. Note that, when interpreting the results, we want the estimated probabilities to be higher than {those drawn from random distributions} for the panels on the top-left to bottom-right diagonal (corresponding to correct identification), and lower than the random ones for the rest of the panels (incorrect identifications).

The first row of Figure~\ref{fig:truth_table_all} clearly indicates that the probability of identifying core galaxies does not depend significantly on galaxy mass or distance to the cluster centre. The average probability of success is $0.67$, clearly showing that we perform much better than random when identifying cluster core galaxies.

The second row, middle panel. shows that the probability of correctly identifying filament galaxies varies from $\sim 0.44$ to $\sim 0.57$. These probabilities are always better than the random chance {($\sim 0.57$)}, but in some cases not by much -- {thereby once again emphasizing that} identifying filament galaxies in the vicinity of massive clusters is difficult, but with large enough samples the statistics will be able to beat the noise. The likelihood of success increases significantly with galaxy mass. The most likely explanation is that filament galaxies are, on average, more massive than field galaxies outside filaments, as previously found in simulations (e.g., \citealt{Veena_18}) and observations (e.g., \citealt{Alpaslan16,Malavasi17,Kraljic18}). These works indicate the galaxy stellar mass increases when getting closer to the spine of the filaments. In our own simulations, the filament galaxy samples contain more massive galaxies than the NCF samples (the median galaxy mass in filaments is $\sim17\%$ higher), enhancing the probability of correct identification at high masses. There is also a small increase in the probability of correctly identifying galaxies when the distance to the cluster centre decreases. This can be explained by the fact that closer to the cluster core the fraction of the volume (and projected area) contained in filaments is larger when compared to that occupied by NCF galaxies. As a result, the fraction of filament galaxies misidentified as NCF due to projection effects is smaller closer to the core. 

The right panel in the middle row shows that we are a bit better  than random at preventing NCF galaxies from contaminating the filament sample at most masses and radial distances, but for the lowest mass galaxies and largest radial distances we fare a bit worse ({by up to $\sim4\%$}). At very large distances from the cluster, the volume occupied by NCF galaxies is so much larger than that occupied by the filaments that projection effects become too strong. When interpreting the properties of filament galaxy samples identified this way it is very important to be aware that in some regions of the parameter space the filament samples will suffer from low purity . 

Finally, the bottom row of Figure~\ref{fig:truth_table_all} completes the picture for the galaxies that are allocated to the NCF category. The probabilities very much mirror what was found for filament galaxies. We are quite successful ({in some cases, 20 percentage points above random}) at distinguishing NCF galaxies from filament ones, although the NCF sample will contain significant contamination from filament galaxies. There is very little radial dependence, but we find some galaxy mass dependence in the opposite sense to the one found for filament galaxies: we are slightly more successful at identifying NCF galaxies at lower galaxy masses, as expected from the discussion above.

\subsection{The dependence of the environmental identification success on cluster mass}
\label{sec:cluster_mass_dependence}
\begin{figure}
    \centering
    \includegraphics[width = 0.45\textwidth]{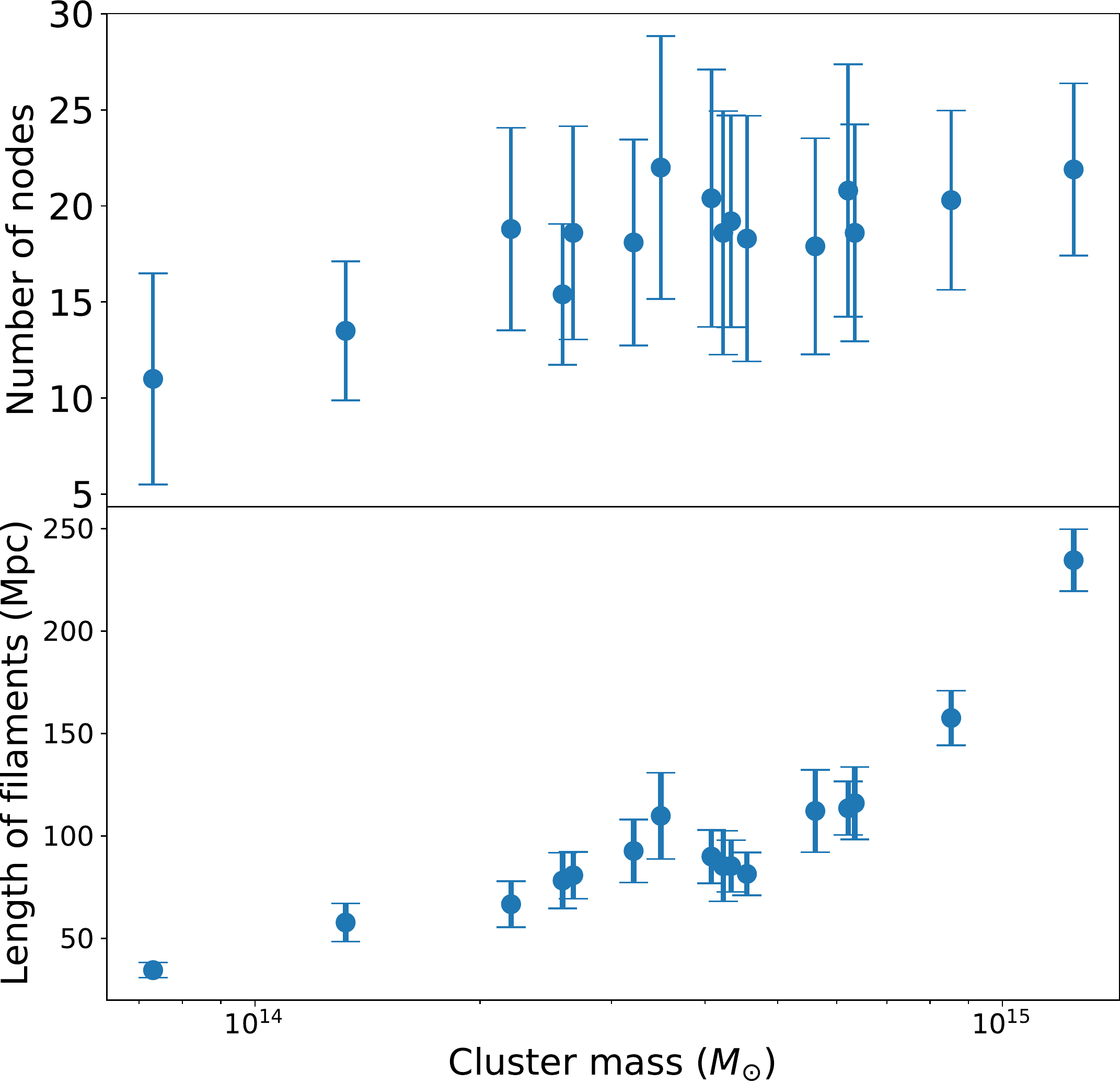}
    \caption{Correlation between filament network properties and the mass of the central cluster for the 160 simulated clusters mass-matched to the WWFCS sample based on  
    \texttt{TheThreeHundred} simulations. The top panel shows the number of nodes in the network as a function of cluster mass. The lower panel shows the total length of all filaments in each network also as a function of cluster mass. Points correspond to the average for the 10 simulated clusters in each mass bin, with the error bars showing the corresponding scatter (standard deviation). {Clear positive correlation are found between cluster mass and the number of nodes or the length of the filament network.}}
    \label{fig:network_properties_vs_cluster_mass}
\end{figure}

In Appendix~\ref{sec:appendix1} (see Figure~\ref{fig:truth_table_SD}) we show that there are significant cluster-to-cluster variations in the probabilities discussed above. This suggests there may be a systematic dependence of our ability to identify galaxy environments on the properties of the central cluster and the filamentary network that surrounds it. The most obvious cluster property to consider is the cluster mass since the galaxy clusters in the WWFCS sample span over 1 order-of-magnitude in mass (from $\sim7\times10^{13}M_\odot$ to over $\sim10^{15}M_\odot$). We consider two properties of the filamentary network that may also have an effect: the total length of all filaments in the network, and number of nodes identified by \texttt{DisPerSE}. These properties encode the complexity and extent of the network, and may therefore influence the probabilities we calculate. Fortunately, both network length and number of nodes correlate reasonably well with cluster mass (Figure~\ref{fig:network_properties_vs_cluster_mass}), and therefore, for the purpose of parameterising the relatively small cluster-to-cluster systematic variations, it suffices to use the cluster mass as main parameter. 

%Figure~\ref{fig:truth_table_SD} shows that there is significant cluster-to-cluster scatter in the calculated probabilities, well above the statistical uncertainty (the standard error for the probability values is between $0.01$ and $0.02$).  This suggest there may be a systematic dependence of our ability to identify galaxy environments on the properties of the central cluster and the filamentary network that surrounds it. The most obvious cluster property to consider is the cluster mass since the galaxy clusters in the WWFCS sample span over 1 order-of-magnitude in mass (from $\sim7\times10^{13}M_\odot$ to over $\sim10^{15}M_\odot$). We consider two properties of the filamentary network that may also have an effect: the total length of all filaments in the network, and number of nodes identified by \texttt{DisPerSE}. These properties encode the complexity and extent of the network, and may therefore influence the probabilities we calculate. Fortunately, both network length and number of nodes correlate reasonably well with cluster mass (Figure~\ref{fig:network_properties_vs_cluster_mass}), and therefore, for the purpose of parameterising the relatively small cluster-to-cluster systematic variations, it suffices to use the cluster mass as main parameter. 

\begin{figure*}
    \centering
    \includegraphics[width = \textwidth]{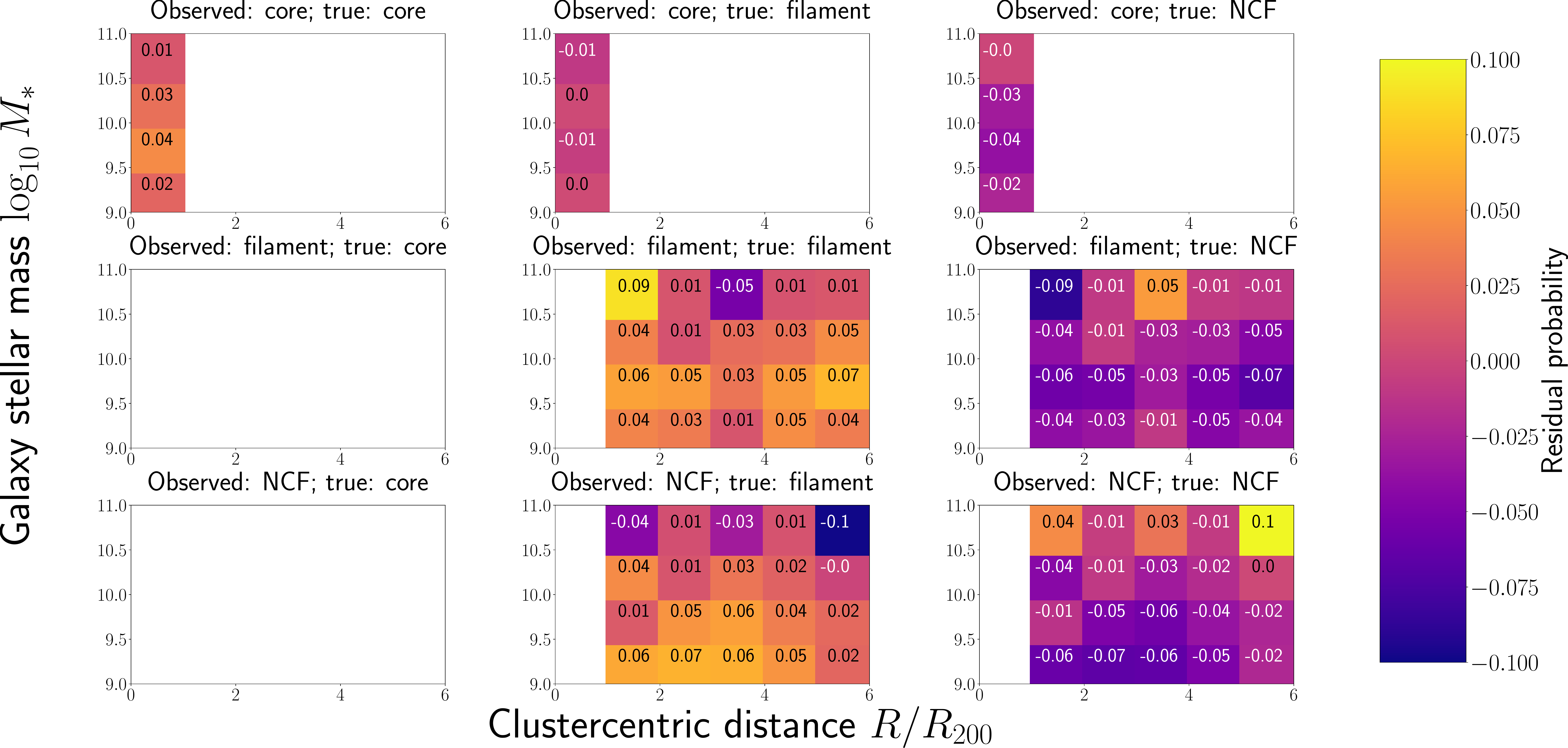}
    \caption{Residual probabilities of the environmental classification showing the dependency on cluster mass. These are calculated as the difference between the average probabilities for the most massive third of the clusters ($M_\text{cluster} > 5 \times 10^{14} M_{\odot}$) and the least massive third ($M_\text{cluster} < 2.5 \times 10^{14} M_{\odot}$). A positive residual indicates that a given probability is higher for the more massive clusters. }
    \label{fig:truth_residual}
\end{figure*}

We divide the clusters into three different mass bins, spanning the full WWFCS mass range: $7.3 \times 10^{13} M_{\odot} < M_{\text{cluster}} < 2.5 \times 10^{14} M_{\odot}$, $2.5 \times 10^{14} M_{\odot} < M_{\text{cluster}} < 5.0 \times 10^{14} M_{\odot}$, and $5.0 \times 10^{14} M_{\odot} < M_{\text{cluster}} < 1.2 \times 10^{15} M_{\odot}$. Each bin contains $\sim50$ simulated clusters. To determine if our success in environmental classification correlates systematically with cluster mass, we calculate the difference in average probabilities between the most massive cluster bin and the least massive one. This is computed for every mass and radial distance bin, and shown in Figure~\ref{fig:truth_residual}. 

For the most massive clusters there is a small but systematic excess in the probability of correctly identifying cluster core galaxies. The difference comes from the fact that the probability of contamination from NCF galaxies is systematically larger for the least massive clusters. This can be understood because the length of the filament network is smaller for clusters with low masses (Figure~\ref{fig:network_properties_vs_cluster_mass}), and therefore the fraction of the volume occupied by NCF galaxies is larger, making projection effects worse.   
The contamination of the core sample from filament galaxies changes very little across cluster masses given the relatively small volume occupied by filaments.  

The second row of Figure~\ref{fig:truth_residual} shows that for high mass clusters we are systematically more successful at identifying filament galaxies correctly, {while still performing generally better than random for lower mass clusters}. This result is statistically significant since the differences in probabilities are generally and systematically larger that the 0.01--0.02 uncertainties. It is likely that the reason for this is simply that in more massive clusters the length of the filament network and thus the fraction of volume occupied by them is larger\footnote{{As a consequence of the correlation between cluster mass and total length of the filament network shown in Figure~\ref{fig:network_properties_vs_cluster_mass} we also find a clear positive correlation between the cluster mass and the fraction of volume occupied by filaments inside  $r<5R_{200}$.}} (Figure~\ref{fig:network_properties_vs_cluster_mass}), implying that the the projection effects leading to the misidentification of filament galaxies are smaller for high mass clusters.

Reciprocally, similar arguments explain how the probability of correctly identifying NCF clusters is higher in low-mass cluster regions (third row of Figure~\ref{fig:truth_residual}). The fractional volume occupied for NCF galaxies increases as the mass of the cluster decreases due to the opposite trend shown by filament galaxies.  

We conclude that relatively small but systematic variations with cluster mass (and correlated filamentary network properties) exist in the probabilities of correctly identifying the environment of galaxies, and it is therefore useful to calculate separate tables for different masses. Given the size of the variations, it suffices to divide the clusters in three mass bins -- further granularity would reduce the statistical accuracy of the calculated probabilities without significantly altering the results. Table~\ref{tab:all} presents in numerical form the probabilities calculates for the complete cluster sample, as shown in Figure~\ref{fig:truth_table_all}. Similarly, Tables~\ref{tab:low}, \ref{tab:med}, and~\ref{tab:high} contain the probabilities for low-, intermediate-, and high-mass clusters.

\section{Conclusions}

Galaxies experience different physical processes in different environments. Next generation wide-field spectroscopic surveys will be able to accurately map out in detail the distribution of galaxies in the cosmic web around galaxy clusters. In \cite{Cornwell_2022} we laid down the framework for developing mock observations to accurately forecast the success in reconstructing cosmic filaments around galaxy clusters for one such survey, the WEAVE wide-field cluster survey (WWFCS). In this paper, we assess the feasibility and accuracy of assigning individual galaxies to different cosmic web environments using a large sample of simulated galaxy clusters from \textsc{TheThreeHundred} project \citep{Cui2018}. In order to do so, we compare the `true' environments we assign to galaxies using the 3D information provided by the simulations with the `observed' environment we assign to the same galaxies using mock observations that take into account the observational constraints and selection effects of the planned WWFCS. We summarize our main findings below.

\begin{enumerate}
    \item Filaments occupy only $\sim 6\%$ of the volume enclosed in a sphere with radius $5R_{200}$ around massive galaxy clusters, but contain $\sim38\%$ of the galaxies with masses above $10^9M_\odot$. {This is calculated using filament thicknesses that decrease with the mass of the main halo (see Sec. \ref{sec:thickness}).} In comparison, galaxies that are neither in the cluster core nor in filaments (NCF) make up $\sim52\%$ of the galaxy population, whilst occupying $93\%$ of the volume. The cluster core itself (defined as the sphere with radius $R_{200}$) contains $\sim 1\%$ of the volume and $\sim 10\%$ of the galaxies. To understand how these different environments affect the properties and evolution of the galaxies that inhabit them we need to be able to associate galaxies to the correct environment and to quantify statistically the uncertainties involved. 
    \item When we allocate galaxies to different environments in the mock observations and compare them to the allocations from the `true' simulations we find that, combining all cluster and galaxy masses, and at all clustercentric distances, we are able to identify core, filament, and NCF galaxies with statistical accuracies (precisions) of 95\% (68\%), 63\% (51\%) and 62\% (68\%) respectively (see definitions in Equations~\ref{eq:accuracy} and~\ref{eq:precission}). This indicates that, while cluster core galaxy samples can be built with a high level of completeness and moderate contamination, the filament and NCF galaxy samples will be significantly contaminated and incomplete due to projection effects, even with good-quality spectroscopic redshifts. 
    \item In our framework, we calculate the probabilities of galaxies being correctly assigned to a given environment, together with the probabilities of misidentifying them as a belonging to a different one (Figure~\ref{fig:truth_table_all} and Table~\ref{tab:all}). We do that as a function of galaxy mass and clustercentric distance. We find that, outside the cluster core (beyond $\sim R_{200}$),  identifying filament galaxies is marginally more successful at high galaxy masses and low clustercentric distances, while the reciprocal is true for NCF galaxies. Generally, the success of the environment allocation is significantly better than random, but sometimes only marginally so. We conclude that identifying the cosmic web environments of galaxies in the vicinity of massive clusters (within a sphere of radius $\sim5R_{200}$ from the cluster centre) is remarkably difficult due to projection effects exacerbated by the magnitude of the galaxies' peculiar velocities (Fingers-of-God).     
    \item We also find that the calculated probabilities vary with the mass of the central cluster and, by association, with properties of the filamentary network such as the number of nodes or the total length of the filaments. We therefore calculate the probabilities for different cluster mass ranges (Tables~\ref{tab:low}, \ref{tab:med}, and~\ref{tab:high}), and find that identifying filament galaxies is marginally more successful around the most massive clusters because their filament networks occupy a relatively larger fraction of the total volume considered. 
\end{enumerate}

We conclude that, in the infall regions surrounding masssive galaxy clusters, associating galaxies with the correct cosmic web environment is highly uncertain. However, applying our statistical framework and probabilities to large spectroscopic samples like the WWFCS will allow us to observationally extract robust and well-defined conclusions on relationships between galaxy properties and their environments.

\section*{Acknowledgements}

DJC, AAS, UK, and MEG acknowledge financial support from the
UK Science and Technology Facilities Council (STFC; through a PhD studentship, and consolidated grant ref
ST/T000171/1). DJC thanks Sean McGee for useful feedback on the contents of this paper. AK is supported by the Ministerio de Ciencia e Innovaci\'{o}n (MICINN) under research grant PID2021-122603NB-C21 and further thanks Dead Can Dance for within the realm of a dying sun.

Funding for the WEAVE facility has been provided by UKRI STFC, the University of Oxford, NOVA, NWO, Instituto de Astrofísica de Canarias (IAC), the Isaac Newton Group partners (STFC, NWO, and Spain, led by the IAC), INAF, CNRS-INSU, the Observatoire de Paris, Région Île-de-France, CONCYT through INAOE, Konkoly Observatory of the Hungarian Academy of Sciences, Max-Planck-Institut für Astronomie (MPIA Heidelberg), Lund University, the Leibniz Institute for Astrophysics Potsdam (AIP), the Swedish Research Council, the European Commission, and the University of Pennsylvania.  The WEAVE Survey Consortium consists of the ING, its three partners, represented by UKRI STFC, NWO, and the IAC, NOVA, INAF, GEPI, INAOE, and individual WEAVE Participants. Please see the relevant footnotes for the WEAVE website\footnote{\url{https://ingconfluence.ing.iac.es/confluence//display/WEAV/The+WEAVE+Project}} and for the full list of granting agencies and grants supporting WEAVE\footnote{\url{https://ingconfluence.ing.iac.es/confluence/display/WEAV/WEAVE+Acknowledgements}}.

This work has been made possible by \texttt{TheThreeHundred} collaboration, which benefits from financial support of the European Union’s Horizon 2020 Research and Innovation programme under the Marie Sk\l{}odowskaw-Curie grant agreement number 734374, i.e. the LACEGAL project. \texttt{TheThreeHundred} simulations used in this paper have been performed in the MareNostrum Supercomputer at the Barcelona Supercomputing Center, thanks to CPU time granted by the Red Espa\~{n}ola de Supercomputaci\'{o}n. 

For the purpose of open access, the authors have applied a creative commons attribution (CC BY) to any journal-accepted manuscript.

%%%%%%%%%%%%%%%%%%%%%%%%%%%%%%%%%%%%%%%%%%%%%%%%%%
\section*{Data Availability}
Data available on request to \texttt{TheThreeHundred} collaboration: https://www.the300-project.org.


\bibliographystyle{mnras}
\bibliography{bibliography} % if your bibtex file is called example.bib

%\begin{thebibliography}{99}
%\end{thebibliography}

%%%%%%%%%%%%%%%%%%%%%%%%%%%%%%%%%%%%%%%%%%%%%%%%%%
\appendix

\section{Cluster-to-cluster probability variation}
\label{sec:appendix1}
\begin{figure*}
    \centering
    \includegraphics[width = \textwidth]{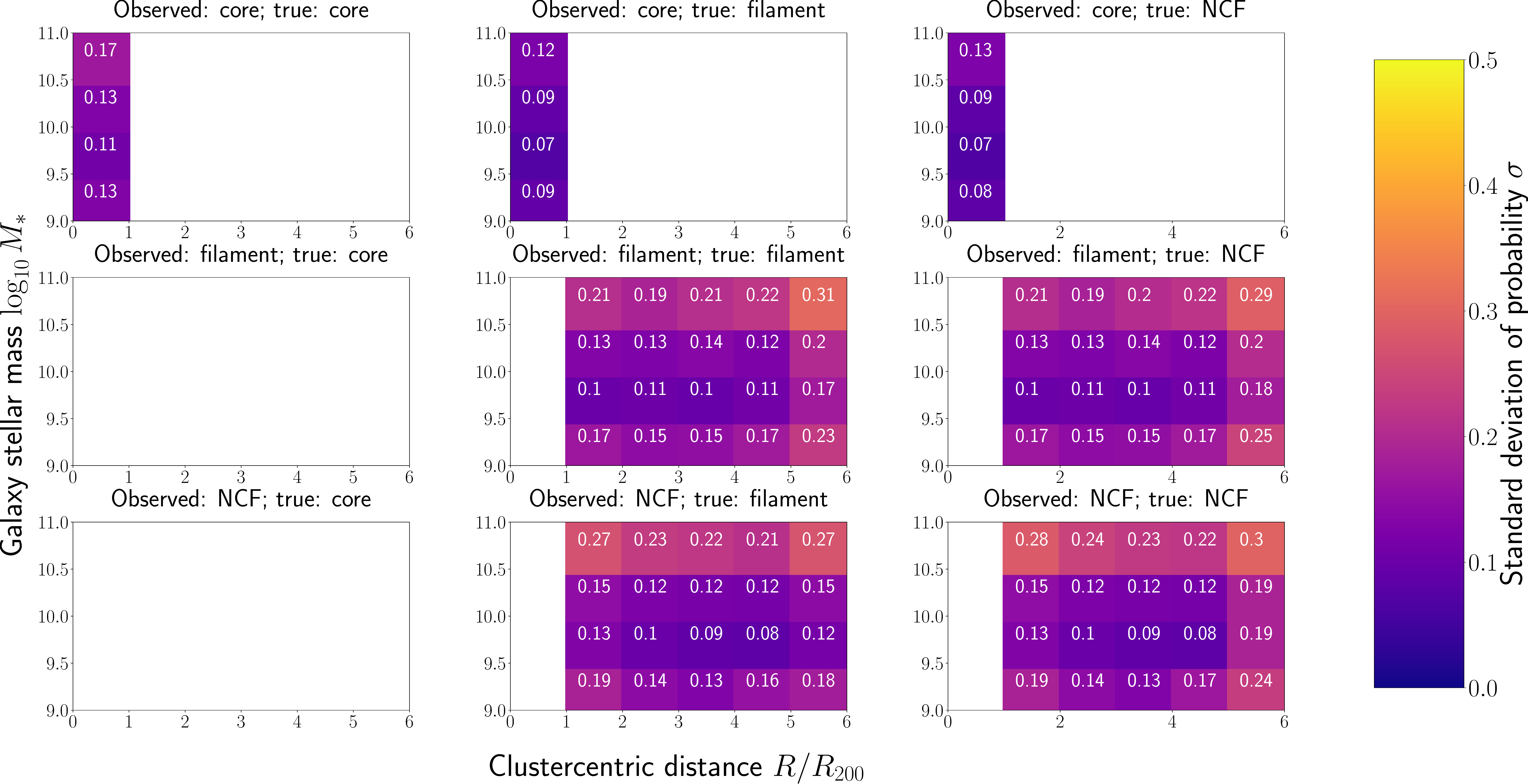}
    \caption{The cluster-to-cluster scatter (standard deviation) of the probabilities shown in Figure~\ref{fig:truth_table_all}.}
    \label{fig:truth_table_SD}
\end{figure*}

To investigate whether there are significant cluster-to-cluster variations in the probabilities of correctly associating galaxies with their environments, we calculate the standard deviation of these probabilities over all clusters and present them in Figure~\ref{fig:truth_table_SD}. The measured scatter is generally well above the statistical uncertainty (the standard error for the probability values is between $0.01$ and $0.02$), demonstrating that there are significant cluster-to-cluster variations. We explore these in Section~\ref{sec:cluster_mass_dependence}.

\begin{table*}
\caption{Probability of galaxies being identified in different cosmic web environments for model clusters of all masses. These probabilities are shown for each environment as a function of galaxy stellar mass and clustercentric distance in a similar arrangement as in Figure~\ref{fig:truth_table_all}. For each mass and distance bin, three numbers are given, each corresponding to each one of the columns in Figure~\ref{fig:truth_table_all}. A number `$-1$' indicates that a given situation is not possible, and corresponds to a white cell in Figure~\ref{fig:truth_table_all}. For instance, we cannot have galaxies classified as `core' at projected radial distances larger than $R_{200}$, and therefore the corresponding probabilities are not defined. Note that the numbers and information presented in this table are exactly the same as those presented in Figure~\ref{fig:truth_table_all}. We provide them here in tabular form for easy access. }
\label{tab:all}
\begin{tabular}{lcccccc}
\hline
Core & $0<r<1R_{200}$ & $1R_{200}<r<2R_{200}$ & $2R_{200}<r<3R_{200}$ & $3R_{200}<r<4R_{200}$ & $4R_{200}<r<5R_{200}$ & $5R_{200}<r<6R_{200}$\\
\hline
$10^{9.0} < M < 10^{9.5}$ & 0.66/0.14/0.20 & $-1$/$-1$/$-1$ & $-1$/$-1$/$-1$ & $-1$/$-1$/$-1$ & $-1$/$-1$/$-1$ & $-1$/$-1$/$-1$ \\
$10^{9.5} < M < 10^{10.0}$ & 0.68/0.14/0.18 & $-1$/$-1$/$-1$ & $-1$/$-1$/$-1$ & $-1$/$-1$/$-1$ & $-1$/$-1$/$-1$ & $-1$/$-1$/$-1$ \\
$10^{10.0} < M < 10^{10.5}$ & 0.65/0.16/0.19 & $-1$/$-1$/$-1$ & $-1$/$-1$/$-1$ & $-1$/$-1$/$-1$ & $-1$/$-1$/$-1$ & $-1$/$-1$/$-1$ \\
$10^{10.5} < M < 10^{11.0}$ & 0.65/0.17/0.18 & $-1$/$-1$/$-1$ & $-1$/$-1$/$-1$ & $-1$/$-1$/$-1$ & $-1$/$-1$/$-1$ & $-1$/$-1$/$-1$ \\
\hline
\hline
Filaments & $0<r<1R_{200}$ & $1R_{200}<r<2R_{200}$ & $2R_{200}<r<3R_{200}$ & $3R_{200}<r<4R_{200}$ & $4R_{200}<r<5R_{200}$ & $5R_{200}<r<6R_{200}$\\
\hline
$10^{9.0} < M < 10^{9.5}$ & $-1$/$-1$/$-1$ & $-1$/0.49/0.51 & $-1$/0.47/0.53 & $-1$/0.48/0.52 & $-1$/0.47/0.53 & $-1$/0.44/0.56 \\
$10^{9.5} < M < 10^{10.0}$ & $-1$/$-1$/$-1$ & $-1$/0.52/0.48 & $-1$/0.53/0.47 & $-1$/0.52/0.48 & $-1$/0.50/0.50 & $-1$/0.50/0.50 \\
$10^{10.0} < M < 10^{10.5}$ & $-1$/$-1$/$-1$ & $-1$/0.54/0.46 & $-1$/0.52/0.48 & $-1$/0.52/0.48 & $-1$/0.53/0.47 & $-1$/0.50/0.50 \\
$10^{10.5} < M < 10^{11.0}$ & $-1$/$-1$/$-1$ & $-1$/0.57/0.43 & $-1$/0.56/0.44 & $-1$/0.57/0.43 & $-1$/0.56/0.44 & $-1$/0.54/0.46 \\
\hline
\hline
NCF & $0<r<1R_{200}$ & $1R_{200}<r<2R_{200}$ & $2R_{200}<r<3R_{200}$ & $3R_{200}<r<4R_{200}$ & $4R_{200}<r<5R_{200}$ & $5R_{200}<r<6R_{200}$\\
\hline
$10^{9.0} < M < 10^{9.5}$ & $-1$/$-1$/$-1$ & $-1$/0.31/0.69 & $-1$/0.29/0.71 & $-1$/0.28/0.72 & $-1$/0.27/0.73 & $-1$/0.29/0.71 \\
$10^{9.5} < M < 10^{10.0}$ & $-1$/$-1$/$-1$ & $-1$/0.33/0.67 & $-1$/0.33/0.67 & $-1$/0.32/0.68 & $-1$/0.31/0.69 & $-1$/0.31/0.69 \\
$10^{10.0} < M < 10^{10.5}$ & $-1$/$-1$/$-1$ & $-1$/0.35/0.65 & $-1$/0.34/0.66 & $-1$/0.34/0.66 & $-1$/0.33/0.67 & $-1$/0.34/0.66 \\
$10^{10.5} < M < 10^{11.0}$ & $-1$/$-1$/$-1$ & $-1$/0.38/0.62 & $-1$/0.41/0.59 & $-1$/0.41/0.59 & $-1$/0.40/0.60 & $-1$/0.40/0.60 \\
\hline
\end{tabular}
\end{table*}

\begin{table*}
\caption{Probability of galaxies being identified in different cosmic web environments for low mass model clusters ($7.3 \times 10^{13} M_{\odot} < M_{\text{cluster}} < 2.5 \times 10^{14} M_{\odot}$). The format is the same as in Table~\ref{tab:all}. }
\label{tab:low}
\begin{tabular}{lcccccc}
\hline
Core & $0<r<1R_{200}$ & $1R_{200}<r<2R_{200}$ & $2R_{200}<r<3R_{200}$ & $3R_{200}<r<4R_{200}$ & $4R_{200}<r<5R_{200}$ & $5R_{200}<r<6R_{200}$\\
\hline
$10^{9.0} < M < 10^{9.5}$ & 0.67/0.13/0.20 & $-1$/$-1$/$-1$ & $-1$/$-1$/$-1$ & $-1$/$-1$/$-1$ & $-1$/$-1$/$-1$ & $-1$/$-1$/$-1$ \\
$10^{9.5} < M < 10^{10.0}$ & 0.65/0.14/0.20 & $-1$/$-1$/$-1$ & $-1$/$-1$/$-1$ & $-1$/$-1$/$-1$ & $-1$/$-1$/$-1$ & $-1$/$-1$/$-1$ \\
$10^{10.0} < M < 10^{10.5}$ & 0.63/0.16/0.21 & $-1$/$-1$/$-1$ & $-1$/$-1$/$-1$ & $-1$/$-1$/$-1$ & $-1$/$-1$/$-1$ & $-1$/$-1$/$-1$ \\
$10^{10.5} < M < 10^{11.0}$ & 0.63/0.18/0.19 & $-1$/$-1$/$-1$ & $-1$/$-1$/$-1$ & $-1$/$-1$/$-1$ & $-1$/$-1$/$-1$ & $-1$/$-1$/$-1$ \\
\hline
\hline
Filaments & $0<r<1R_{200}$ & $1R_{200}<r<2R_{200}$ & $2R_{200}<r<3R_{200}$ & $3R_{200}<r<4R_{200}$ & $4R_{200}<r<5R_{200}$ & $5R_{200}<r<6R_{200}$\\
\hline
$10^{9.0} < M < 10^{9.5}$ & $-1$/$-1$/$-1$ & $-1$/0.48/0.52 & $-1$/0.47/0.53 & $-1$/0.47/0.53 & $-1$/0.45/0.55 & $-1$/0.43/0.57 \\
$10^{9.5} < M < 10^{10.0}$ & $-1$/$-1$/$-1$ & $-1$/0.49/0.51 & $-1$/0.50/0.50 & $-1$/0.50/0.50 & $-1$/0.48/0.52 & $-1$/0.46/0.54 \\
$10^{10.0} < M < 10^{10.5}$ & $-1$/$-1$/$-1$ & $-1$/0.52/0.48 & $-1$/0.53/0.47 & $-1$/0.51/0.49 & $-1$/0.53/0.47 & $-1$/0.48/0.52 \\
$10^{10.5} < M < 10^{11.0}$ & $-1$/$-1$/$-1$ & $-1$/0.51/0.49 & $-1$/0.57/0.43 & $-1$/0.63/0.37 & $-1$/0.57/0.43 & $-1$/0.55/0.45 \\
\hline
\hline
NCF & $0<r<1R_{200}$ & $1R_{200}<r<2R_{200}$ & $2R_{200}<r<3R_{200}$ & $3R_{200}<r<4R_{200}$ & $4R_{200}<r<5R_{200}$ & $5R_{200}<r<6R_{200}$\\
\hline
$10^{9.0} < M < 10^{9.5}$ & $-1$/$-1$/$-1$ & $-1$/0.29/0.71 & $-1$/0.26/0.74 & $-1$/0.26/0.74 & $-1$/0.26/0.74 & $-1$/0.28/0.72 \\
$10^{9.5} < M < 10^{10.0}$ & $-1$/$-1$/$-1$ & $-1$/0.32/0.68 & $-1$/0.29/0.71 & $-1$/0.29/0.71 & $-1$/0.29/0.71 & $-1$/0.30/0.70 \\
$10^{10.0} < M < 10^{10.5}$ & $-1$/$-1$/$-1$ & $-1$/0.33/0.67 & $-1$/0.35/0.65 & $-1$/0.33/0.67 & $-1$/0.33/0.67 & $-1$/0.33/0.67 \\
$10^{10.5} < M < 10^{11.0}$ & $-1$/$-1$/$-1$ & $-1$/0.42/0.58 & $-1$/0.42/0.58 & $-1$/0.45/0.55 & $-1$/0.39/0.61 & $-1$/0.50/0.50 \\
\hline
\end{tabular}
\end{table*}
\begin{table*}
\caption{Probability of galaxies being identified in different cosmic web environments for intermediate mass model clusters ($2.5 \times 10^{14} M_{\odot} < M_{\text{cluster}} < 5.0 \times 10^{14} M_{\odot}$). The format is the same as in Table~\ref{tab:all}.}
\label{tab:med}
\begin{tabular}{lcccccc}
\hline
Core & $0<r<1R_{200}$ & $1R_{200}<r<2R_{200}$ & $2R_{200}<r<3R_{200}$ & $3R_{200}<r<4R_{200}$ & $4R_{200}<r<5R_{200}$ & $5R_{200}<r<6R_{200}$\\
\hline
$10^{9.0} < M < 10^{9.5}$ & 0.64/0.15/0.21 & $-1$/$-1$/$-1$ & $-1$/$-1$/$-1$ & $-1$/$-1$/$-1$ & $-1$/$-1$/$-1$ & $-1$/$-1$/$-1$ \\
$10^{9.5} < M < 10^{10.0}$ & 0.68/0.13/0.19 & $-1$/$-1$/$-1$ & $-1$/$-1$/$-1$ & $-1$/$-1$/$-1$ & $-1$/$-1$/$-1$ & $-1$/$-1$/$-1$ \\
$10^{10.0} < M < 10^{10.5}$ & 0.64/0.15/0.21 & $-1$/$-1$/$-1$ & $-1$/$-1$/$-1$ & $-1$/$-1$/$-1$ & $-1$/$-1$/$-1$ & $-1$/$-1$/$-1$ \\
$10^{10.5} < M < 10^{11.0}$ & 0.66/0.16/0.17 & $-1$/$-1$/$-1$ & $-1$/$-1$/$-1$ & $-1$/$-1$/$-1LaTeX Warning: Reference `firstpage' on page 1 undefined on input line 125.
$ & $-1$/$-1$/$-1$ & $-1$/$-1$/$-1$ \\
\hline
\hline
Filaments & $0<r<1R_{200}$ & $1R_{200}<r<2R_{200}$ & $2R_{200}<r<3R_{200}$ & $3R_{200}<r<4R_{200}$ & $4R_{200}<r<5R_{200}$ & $5R_{200}<r<6R_{200}$\\
\hline
$10^{9.0} < M < 10^{9.5}$ & $-1$/$-1$/$-1$ & $-1$/0.49/0.51 & $-1$/0.47/0.53 & $-1$/0.49/0.51 & $-1$/0.49/0.51 & $-1$/0.44/0.56 \\
$10^{9.5} < M < 10^{10.0}$ & $-1$/$-1$/$-1$ & $-1$/0.50/0.50 & $-1$/0.51/0.49 & $-1$/0.51/0.49 & $-1$/0.48/0.52 & $-1$/0.49/0.51 \\
$10^{10.0} < M < 10^{10.5}$ & $-1$/$-1$/$-1$ & $-1$/0.51/0.49 & $-1$/0.50/0.50 & $-1$/0.49/0.51 & $-1$/0.49/0.51 & $-1$/0.47/0.53 \\
$10^{10.5} < M < 10^{11.0}$ & $-1$/$-1$/$-1$ & $-1$/0.53/0.47 & $-1$/0.52/0.48 & $-1$/0.55/0.45 & $-1$/0.54/0.46 & $-1$/0.51/0.49 \\
\hline
\hline
NCF & $0<r<1R_{200}$ & $1R_{200}<r<2R_{200}$ & $2R_{200}<r<3R_{200}$ & $3R_{200}<r<4R_{200}$ & $4R_{200}<r<5R_{200}$ & $5R_{200}<r<6R_{200}$\\
\hline
$10^{9.0} < M < 10^{9.5}$ & $-1$/$-1$/$-1$ & $-1$/0.33/0.67 & $-1$/0.32/0.68 & $-1$/0.33/0.67 & $-1$/0.27/0.73 & $-1$/0.30/0.70 \\
$10^{9.5} < M < 10^{10.0}$ & $-1$/$-1$/$-1$ & $-1$/0.32/0.68 & $-1$/0.33/0.67 & $-1$/0.31/0.69 & $-1$/0.31/0.69 & $-1$/0.31/0.69 \\
$10^{10.0} < M < 10^{10.5}$ & $-1$/$-1$/$-1$ & $-1$/0.32/0.68 & $-1$/0.31/0.69 & $-1$/0.31/0.69 & $-1$/0.31/0.69 & $-1$/0.36/0.64 \\
$10^{10.5} < M < 10^{11.0}$ & $-1$/$-1$/$-1$ & $-1$/0.39/0.61 & $-1$/0.37/0.63 & $-1$/0.38/0.62 & $-1$/0.39/0.61 & $-1$/0.36/0.64 \\
\hline
\end{tabular}
\end{table*}

\begin{table*}
\caption{Probability of galaxies being identified in different cosmic web environments for high mass model clusters ($5.0 \times 10^{14} M_{\odot} < M_{\text{cluster}} < 1.2 \times 10^{15} M_{\odot}$). The format is the same as in Table~\ref{tab:all}.}
\label{tab:high}
\begin{tabular}{lcccccc}
\hline
Core & $0<r<1R_{200}$ & $1R_{200}<r<2R_{200}$ & $2R_{200}<r<3R_{200}$ & $3R_{200}<r<4R_{200}$ & $4R_{200}<r<5R_{200}$ & $5R_{200}<r<6R_{200}$\\
\hline
$10^{9.0} < M < 10^{9.5}$ & 0.69/0.14/0.18 & $-1$/$-1$/$-1$ & $-1$/$-1$/$-1$ & $-1$/$-1$/$-1$ & $-1$/$-1$/$-1$ & $-1$/$-1$/$-1$ \\
$10^{9.5} < M < 10^{10.0}$ & 0.70/0.14/0.17 & $-1$/$-1$/$-1$ & $-1$/$-1$/$-1$ & $-1$/$-1$/$-1$ & $-1$/$-1$/$-1$ & $-1$/$-1$/$-1$ \\
$10^{10.0} < M < 10^{10.5}$ & 0.66/0.17/0.18 & $-1$/$-1$/$-1$ & $-1$/$-1$/$-1$ & $-1$/$-1$/$-1$ & $-1$/$-1$/$-1$ & $-1$/$-1$/$-1$ \\
$10^{10.5} < M < 10^{11.0}$ & 0.64/0.17/0.19 & $-1$/$-1$/$-1$ & $-1$/$-1$/$-1$ & $-1$/$-1$/$-1$ & $-1$/$-1$/$-1$ & $-1$/$-1$/$-1$ \\
\hline
\hline
Filaments & $0<r<1R_{200}$ & $1R_{200}<r<2R_{200}$ & $2R_{200}<r<3R_{200}$ & $3R_{200}<r<4R_{200}$ & $4R_{200}<r<5R_{200}$ & $5R_{200}<r<6R_{200}$\\
\hline
$10^{9.0} < M < 10^{9.5}$ & $-1$/$-1$/$-1$ & $-1$/0.52/0.48 & $-1$/0.50/0.50 & $-1$/0.48/0.52 & $-1$/0.50/0.50 & $-1$/0.47/0.53 \\
$10^{9.5} < M < 10^{10.0}$ & $-1$/$-1$/$-1$ & $-1$/0.55/0.45 & $-1$/0.55/0.45 & $-1$/0.53/0.47 & $-1$/0.54/0.46 & $-1$/0.53/0.47 \\
$10^{10.0} < M < 10^{10.5}$ & $-1$/$-1$/$-1$ & $-1$/0.56/0.44 & $-1$/0.53/0.47 & $-1$/0.54/0.46 & $-1$/0.56/0.44 & $-1$/0.53/0.47 \\
$10^{10.5} < M < 10^{11.0}$ & $-1$/$-1$/$-1$ & $-1$/0.60/0.40 & $-1$/0.58/0.42 & $-1$/0.58/0.42 & $-1$/0.58/0.42 & $-1$/0.56/0.44 \\
\hline
\hline
NCF & $0<r<1R_{200}$ & $1R_{200}<r<2R_{200}$ & $2R_{200}<r<3R_{200}$ & $3R_{200}<r<4R_{200}$ & $4R_{200}<r<5R_{200}$ & $5R_{200}<r<6R_{200}$\\
\hline
$10^{9.0} < M < 10^{9.5}$ & $-1$/$-1$/$-1$ & $-1$/0.35/0.65 & $-1$/0.33/0.67 & $-1$/0.32/0.68 & $-1$/0.31/0.69 & $-1$/0.31/0.69 \\
$10^{9.5} < M < 10^{10.0}$ & $-1$/$-1$/$-1$ & $-1$/0.34/0.66 & $-1$/0.34/0.66 & $-1$/0.35/0.65 & $-1$/0.33/0.67 & $-1$/0.33/0.67 \\
$10^{10.0} < M < 10^{10.5}$ & $-1$/$-1$/$-1$ & $-1$/0.37/0.63 & $-1$/0.36/0.64 & $-1$/0.36/0.64 & $-1$/0.34/0.66 & $-1$/0.33/0.67 \\
$10^{10.5} < M < 10^{11.0}$ & $-1$/$-1$/$-1$ & $-1$/0.37/0.63 & $-1$/0.43/0.57 & $-1$/0.42/0.58 & $-1$/0.40/0.60 & $-1$/0.40/0.60 \\
\hline
\end{tabular}
\end{table*}

%%%%%%%%%%%%%%%%% APPENDICES %%%%%%%%%%%%%%%%%%%%%

%%%%%%%%%%%%%%%%%%%%%%%%%%%%%%%%%%%%%%%%%%%%%%%%%%

% Don't change these lines
\bsp	% typesetting comment
\label{lastpage}
\end{document}